\documentclass{iopart}
\usepackage{setstack}
\usepackage{iopams}
\usepackage{graphicx}

\begin{document}

\title{Searching for gravitational waves from Cassiopeia A with LIGO}

\newcounter{affilcount}
\newcommand{\makeaffil}[1]{\addtocounter{affilcount}{1}\edef#1{$^{\arabic{affilcount}}$}}
\newcommand{\affand}{$^,$}

\makeaffil{\AG}
\makeaffil{\AH}
\makeaffil{\AN}
\makeaffil{\CL}
\makeaffil{\CT}
\makeaffil{\LO}
\makeaffil{\PU}
\makeaffil{\TC}
\makeaffil{\BB}
\makeaffil{\GU}
\makeaffil{\MU}
\makeaffil{\SH}
\makeaffil{\UW}

\author{
K~Wette\AN,
B~J~Owen\PU,
B~Allen\AH\affand\UW,
M~Ashley\AN,
J~Betzwieser\CT,
N~Christensen\CL,
T~D~Creighton\TC,
V~Dergachev\MU,
I~Gholami\AG,
E~Goetz\MU,
R~Gustafson\MU,
D~Hammer\UW,
D~I~Jones\SH,
B~Krishnan\AG,
M~Landry\LO,
B~Machenschalk\AH,
D~E~McClelland\AN,
G~Mendell\LO,
C~J~Messenger\AH,
M~A~Papa\AG\affand\UW,
P~Patel\CT,
M~Pitkin\GU,
H~J~Pletsch\AH,
R~Prix\AH,
K~Riles\MU,
L~Sancho~de~la~Jordana\BB,
S~M~Scott\AN,
A~M~Sintes\BB\affand\AG,
M~Trias\BB,
J~T~Whelan\AG~and
G~Woan\GU
}
\address{\AG Albert-Einstein-Institut, Max-Planck-Institut f\"ur Gravitationsphysik, D-14476 Golm, Germany}
\address{\AH Albert-Einstein-Institut, Max-Planck-Institut f\"ur Gravitationsphysik, D-30167 Hannover, Germany}
\address{\AN Australian National University, Canberra, 0200, Australia}
\address{\CL Carleton College, Northfield, MN  55057, USA}
\address{\CT LIGO - California Institute of Technology, Pasadena, CA  91125, USA}
\address{\LO LIGO Hanford Observatory, Richland, WA  99352, USA}
\address{\PU The Pennsylvania State University, University Park, PA  16802, USA}
\address{\TC The University of Texas at Brownsville and Texas Southmost College, Brownsville, TX  78520, USA}
\address{\BB Universitat de les Illes Balears, E-07122 Palma de Mallorca, Spain}
\address{\GU University of Glasgow, Glasgow, G12 8QQ, United Kingdom}
\address{\MU University of Michigan, Ann Arbor, MI  48109, USA}
\address{\SH University of Southampton, Southampton, SO17 1BJ, United Kingdom}
\address{\UW University of Wisconsin-Milwaukee, Milwaukee, WI  53201, USA}

\ead{karl.wette@ligo.org}

\begin{abstract}
We describe a search underway for periodic gravitational waves from the central
compact object in the supernova remnant Cassiopeia A. The object is the youngest
likely neutron star in the Galaxy. Its position is well known, but the object
does not pulse in any electromagnetic radiation band and thus presents a
challenge in searching the parameter space of frequency and frequency
derivatives. We estimate that a fully coherent search can, with a
reasonable amount of time on a computing cluster, achieve a sensitivity at
which it is theoretically possible (though not likely) to observe a signal even
with the initial LIGO noise spectrum. Cassiopeia A is only the second object
after the Crab pulsar for which this is true.
The search method described here can also obtain interesting results for
similar objects with current LIGO sensitivity.
\end{abstract}

\maketitle

\section{Introduction}

The LIGO Scientific Collaboration (LSC) has so far published three types of
searches for periodic gravitational waves (GWs): searches for known
non-accreting pulsars \cite{Abbott:2003yq, Abbott:2004ig, Abbott:2007ce,
Abbott:2008fx}, for
the non-pulsing low-mass X-ray binary Sco X-1 \cite{Abbott:2006vg,
Abbott:2007tw}, and all-sky searches for as yet unknown neutron stars
\cite{Abbott:2006vg, Abbott:2005pu, Abbott:2007tda, Abbott:2008uq}.
The first and last types of search are approaching the indirect upper limits on
gravitational wave emission inferred from the observed spindowns (spin
frequency derivatives) of pulsars and supernova-based estimates of the neutron
star population of the galaxy \cite{Abbott:2006vg}.

Here we discuss the first of a fourth type of search for periodic gravitational
waves: directed searches, which target likely neutron stars whose sky position
is known to high accuracy, but whose spin frequencies and frequency evolution are
not known at all.
We describe such a search, which is currently underway, directed at
the central compact object in the supernova remnant Cassiopeia A (Cas A).
The data analysis
challenge is to search a large parameter space of possible frequencies and
frequency evolutions. We describe the object, estimate the computational costs
of the search, and show that when the search of data from LIGO's recently
completed S5 run is completed, it will beat the indirect limit on GW strain for
Cas A.
We also indicate how cost and sensitivity estimates can be extended to
other directed searches.

\section{The central compact object in Cas A}

Cas A is a core-collapse supernova remnant, currently the youngest known in the
Galaxy \cite{fesen:age}. A central X-ray point source was discovered in
first-light images taken by the Chandra X-Ray Observatory, indicating the
presence of a central compact object (CCO). The nature of the CCO remains
uncertain. No radio pulsations or $\gamma$-ray emission have been observed,
and there is no pulsar wind nebula observed in X-ray or radio; it is unlikely
therefore that the CCO is an active pulsar \cite{fesen:xps}.
Proposed explanations
include that it might be a young radio-quiet neutron star, or an accretion disk
associated with a neutron star or black hole, or that it might be related to a
type of slowly rotating neutron star known as an anomalous X-ray pulsar (AXP)
or a soft $\gamma$-ray repeater (SGR) \cite{fesen:xps, chakrabarty:xps}.
Only in the first scenario could GW emission be detectable by LIGO.
What makes Cas A an attractive target is its youth:
the stars with the highest indirect limits (see next section) on gravitational radiation
are young, and one could argue on theoretical grounds that any
deformations left over from the violent birth of the star have had less time to
be smoothed away by mechanisms such as viscoelastic creep.
Young stars also spin more quickly than old ones.
Of the seven confirmed CCOs, only two (possibly three) have measured spin
periods~\cite{DeLuca:2007cq}.
The fastest is radiating gravitational waves at 20~Hz, just below the LIGO
frequency band, but the other CCOs are also much older than Cas A.

For the purpose of a directed search, we need to know the object's right
ascension and declination.
Chandra observations \cite{fesen:xps} have obtained these to sub-arcsecond
accuracy [$\alpha = 23{\rm h}\, 23{\rm m}\, (27.945 \pm 0.05){\rm s}$, $\delta =
58^\circ 48' (42.51 \pm 0.4)''$], which is sufficient for any GW observation.
In order to
define the range of search parameters and give an indirect limit on GW emission
from the object, we also need the distance, age, and moment of inertia. The
distance to Cas A has been estimated from the radial velocities of knots of
ejected material to be $3.4^{+0.3}_{-0.1}$ kpc \cite{reed:distance}. Extrapolation of
the proper motions of outer ejecta knots suggest a convergence date of $1681
\pm 19$, consistent with a possible observation by John Flamsteed in 1680
\cite{fesen:age}.
Since computational costs are higher for younger objects, we play it safe by
taking 300 years (the approximate lower bound) as our fiducial age estimate.
In what
follows we use the canonical neutron star moment of inertia of
$10^{45}$~g~cm$^2$, although modern equations of state predict values higher
for most neutron stars by a factor 2 or 3 \cite{Bejger:2005jy}.

\section{Indirect limits}

Indirect limits on the gravitational wave emission from rotating neutron
stars are found by assuming that the gravitational wave luminosity
is bounded by the time derivative of the total rotational kinetic energy:
\begin{equation}
\left( \frac{dE}{dt} \right)_{\rm gw} =
\frac{32 G}{5 c^5} I_{zz}^2 \epsilon^2 (\pi f)^6
\le
-\frac{d}{dt} \left( \frac{1}{2} \pi^2 I_{zz} f^2 \right)
= -\left( \frac{dE}{dt} \right)_{\rm rot} ,
\end{equation}
where $\epsilon$ is the equatorial ellipticity, $I_{zz}$ the principal moment
of inertia (assumed constant), and $f$ the gravitational wave frequency
(assumed to be twice the spin frequency) \cite{Abbott:2006vg, PhysRevD.20.351}.
This condition is rearranged to give the ``spindown'' upper bounds on the
ellipticity and the GW strain tensor amplitude $h_0$:
\begin{equation}
\label{spindown limits}
\epsilon \le \sqrt{\frac{5 c^5}{32 \pi^4 G I_{zz}} \frac{-\dot{f}}{f^5}} \;,
\qquad
h_0 \le \frac{1}{D} \sqrt{\frac{5 G I_{zz}}{2 c^3} \frac{-\dot{f}}{f}} \;,
\end{equation}
The second limit is found from the first by substituting
\begin{equation}
h_0 = \frac{4 \pi^2 G}{c^4} \frac{I_{zz} \epsilon f^2}{D} \; .
\end{equation}
where $D$ is the distance of the source \cite{Abbott:2006vg, jks}.

For a directed search, the GW frequency $f$ and its time derivative
$\dot{f}$ are unknown, but the age is known.
If we assume that the star is spinning down with $\dot{f} \propto f^n$, and
that it is currently spinning significantly more slowly than it was at birth,
we can relate the frequency evolution to the characteristic age $\tau$ and
braking index $n$ by \cite{ostriker&gunn,
palomba, Abbott:2006vg}
\begin{equation}
\tau \approx \frac{1}{n - 1} \left(\frac{f}{-\dot{f}}\right) \;,
\qquad
n = \frac{f\ddot{f}}{\dot{f}^2}.
\end{equation}
If the spindown is dominated by GW from a constant mass quadrupole, then
$n=5$ and $\tau$ is the true age of the star.
Substituting into the spindown limits \eref{spindown limits} gives
\begin{equation}
\epsilon_{\rm age} \le \sqrt{\frac{5 c^5}{128 \pi^4 G I_{zz} \tau f^4}} \;,
\qquad
h_{\rm age} \le \frac{1}{D} \sqrt{\frac{5 G I_{zz}}{8 c^3 \tau}} \;.
\end{equation}
Using the numbers for Cas A from the previous section we get
\begin{eqnarray}
h_{\rm age} &\le 1.2 \times 10^{-24} \left(\frac{3.4 \mbox{ kpc}}{D}\right)
\sqrt{\left(\frac{I_{zz}}{10^{45} \mbox{ g cm}^2}\right) \left(\frac{300 \mbox{
years}}{\tau}\right)}
\\
\epsilon_{\rm age} &\le 3.9 \times 10^{-4} \left(\frac{100 \mbox{
Hz}}{f}\right)^2 \sqrt{\left(\frac{10^{45} \mbox{ g cm}^2}{I_{zz}}\right)
\left(\frac{300 \mbox{ years}}{\tau}\right)} .
\end{eqnarray}
Below we will consider searches over the range $n=2$--7, including the
possibility that $n$ has changed since the supernova and thus a
lifetime-averaged value is appropriate.
Considering  this, the uncertainty in $D$, and the fact that $I_{zz}$ may
be triple our fiducial value (see discussion in~\cite{Abbott:2007ce}),
these fiducial indirect upper limits are uncertain by about a factor of 2.
Some theories of quark matter allow for ellipticities in the range
indicated, though normal neutron star models do not \cite{Owen:2005fn,
Lin:2007rz, Haskell:2007zz}.
An internal magnetic field of order $10^{16}$~G could also produce such
ellipticities \cite{Cutler:2002nw, Owen:2006tz, Haskell:2007bh, Akgun:2007ph},
although it is not clear if such a field is stable, and if the external field
is this strong then the star by now has spun down out of the LIGO frequency
band.
The age-based indirect limits serve, like the spindown limits, as indicators of
which objects are interesting, but since they are based on less information
they are not as solid as the spindown limits.
It is not known if Cas A spins in the LIGO band (period $\le50$~ms), and indeed
only 10\% of known pulsars do so \cite{atnf}.
Thus a search such as we describe could detect an object on the speculative end
of the range of theoretical predictions.

\section{Search method}

The LSC uses both fully coherent \cite{Abbott:2003yq, Abbott:2004ig,
Abbott:2007ce, Abbott:2008fx, Abbott:2006vg} and
semi-coherent \cite{Abbott:2007tw, Abbott:2005pu, Abbott:2007tda,
Abbott:2008uq} methods to search for periodic gravitational waves.
Semi-coherent methods are computationally cheaper than coherent methods, but
coherent methods can achieve greater sensitivity if the cost is feasible.

For a young neutron star such as Cas A the integration time needed is short
enough (see next section) for us to pursue enhanced sensitivity without
undue computational cost.
We therefore use the fully coherent $\mathcal{F}$-statistic search
\cite{Abbott:2006vg}, as implemented by the \texttt{ComputeFStatistic\_v2}
routine in the LSC Algorithm Library \cite{LAL}.
This routine computes optimal filters for the gravitational wave signal,
including modulation by the detector beam patterns, in multiple interferometers
which are treated as a coherent network \cite{jks, Cutler:2005hc}.
This search uses data from the 4km LIGO interferometers at Hanford, WA,
and Livingston, LA.

The computation is conducted in the frequency domain
using short Fourier transforms (SFTs) of segments of strain data,
typically of 30 minutes duration so that the GW
frequency will remain in one frequency bin over the length of the SFT
\cite{Abbott:2006vg}.
The SFTs are vetoed by a suite of data quality flags to remove poorer
quality data.
For windows of up to 15 days during the first year of the S5 run the duty
cycle -- the ratio of post-veto SFT live time to total time span, averaged
over interferometers -- can somewhat exceed 70\%.

A search for a young neutron star such as Cas A, which is younger than
objects considered in previous LIGO multi-template searches, must cover a
greater spindown parameter space including a second frequency derivative
(see next section).
This has required the extension of existing LSC software to efficiently cover a
three-dimensional space using the parameter space metric. The points are
distributed on a body-centered cubic (bcc or $A_3^*$) lattice, which is known
to be the optimal lattice covering in three dimensions \cite{conway}.

In the event no plausible signal is found, we will set upper limits by methods
similar to the frequentist analyses in \cite{Abbott:2003yq, Abbott:2006vg}.
These are based on Monte Carlo simulations searching the data for a multitude of
software-injected signals with a distribution of amplitudes, inclination
angles, and polarization angles in each frequency bin.
We will also test on a smaller set of simulated signals which were hardware
injected into the S5 data.

\section{Estimated cost and sensitivity}

The sensitivity of a search for periodic signals can be put in terms of the
95\% confidence limit on GW strain tensor amplitude, which takes the form
\begin{equation}
\label{h95}
h_0^{95\%} = \Theta \sqrt{S_h(f)/T_\mathrm{dat}}.
\end{equation}
Here $S_h$ is the strain noise power spectral density, $T_\mathrm{dat}$ is the
data live time, and $\Theta$ is a statistical threshold factor which depends on
the parameter space and other details of the data analysis pipeline.
For a coherent multi-interferometer search, the limits add in inverse
quadrature.
Monte Carlo simulations searching for injected signals from Cas A, as well
as the results of the similar multi-template Crab search
\cite{Abbott:2008fx}, indicate that
$\Theta$ is in the mid-30s for a directed search, and thus we use 35 in
our estimates below.
Because $\Theta$ is determined by the tail of a Gaussian distribution, it is
very weakly dependent on the volume of parameter space searched.
However the data live time $T_\mathrm{dat}$ is computationally limited and thus
does depend on the parameter space.

The parameter space range is chosen as follows.
The frequency band is chosen to be 100--300~Hz, which surrounds the band
where the LIGO interferometers are most sensitive.
As we shall see below, this is roughly the band over which a directed
search can beat the indirect limit on $h_0$ with reasonable computational
cost.
The frequency derivative ranges are chosen based on considering braking indices
$n$ in the range 2--7.
This range covers all known pulsars, except the Vela pulsar which is visibly
interacting with its wind nebula (nonexistent for Cas A).
It also includes the values for radiation dominated by a static dipole or
quadrupole ($n = 3$ or $5$) as well as a saturated $r$-mode ($n = 7$) \cite{Owen:1998xg}.
Thus the range of each frequency derivative depends on the lower derivatives,
and we have
\begin{equation}
\label{limits}
100\mbox{ Hz} \le f \le 300\mbox{ Hz},
\quad
\frac{f}{6\tau} \le -\dot{f} \le \frac{f}{\tau},
\quad
\frac{2\dot{f}^2}{f} \le \ddot{f} \le \frac{7\dot{f}^2}{f}.
\end{equation}
Note that the range of $\ddot{f}$ by definition is related to the
present-day braking index, while the range of $\dot{f}$ corresponds to an
average braking index over the lifetime of the star.
Thus we allow for the braking index varying over time between the indicated
limits.

There remains the problem of efficiently tiling, or choosing specific points in
parameter space for which to compute the $\mathcal{F}$-statistic.
It is straightforward to apply the method of \cite{Owen:1995tm} to find the
parameter space metric \cite{whitbeck}
\begin{equation}
\gamma_{jk} = \frac{4\pi^2 T_\mathrm{span}^{j+k+2} (j+1)(k+1)}
{(j+2)!(k+2)!(j+k+3)},
\end{equation}
where the components are with respect to the $k$th derivative of the GW
frequency at the beginning of the
observation, $T_\mathrm{span}$ is the total duration of data (including
dropouts),
and the indices $j, k$ take integer values between 0 and the highest
derivative considered (2 for Cas A).
This metric, which is the Fisher information matrix with a phase constant
projected out, is used to set up an efficient tiling which takes advantage of
the covariances between parameters.
The number of points needed for an optimal (bcc or $A_3^*$) tiling is given by
\cite{Prix:2007ks}
\begin{equation}
\label{Np}
N_p \simeq 0.19 \mu^{-3/2} \sqrt{\det\gamma} \;\frac{f_{\max}^3}{\tau^3},
\end{equation}
where $\mu$ is the mismatch and we have performed the integral in
equation~(24) of \cite{Prix:2007ks} using the ranges \eref{limits} and
discarding the lower bound on frequency, which is only a few percent
correction.
We determine the highest frequency derivative needed by finding $k$ such that
$\gamma_{kk}\Delta_k^2 > \mu$, where $\Delta_k$ is the range of the $k$th
frequency derivative and we take $\mu$ to be 20\% (typical for periodic signal
searches).
In our case $\ddot{f}$ is required for $T_\mathrm{span}$ greater than about a
week; as shown below, this applies for any search competitive with the
indirect limit.

Since equation~\eref{Np} is obtained by dividing the proper volume of the
parameter space by the proper volume per template, we expect it to
underestimate $N_p$ of a practical implementation due to the need to
cover the edges of the parameter space.
Because the extent of our parameter space in $\ddot{f}$ is often comparable
to or less than the unit cell length of a single template, we expect that
an ideal lattice covering would require several times the ideal number of
templates in \eref{Np}.
Technical limitations of a speedy---and therefore simple---template bank
generation algorithm also require us to lay extra templates to guarantee
that the edges of the parameter space are completely covered.
We have found from Monte Carlo simulations that the combination of these
effects can cause \eref{Np} to underestimate $N_p$ by up to an order of
magnitude.
Even in this worst case, without any improvement of existing template bank
algorithms, the computational cost is still feasible since our fiducial
estimate below is for a small number of computing nodes.
The size of the template bank should not significantly affect the upper
limits, which are very weakly dependent on the number of templates and thus
on the number of statistical trials.

\begin{figure}
\vspace*{30pt}
\centerline{\includegraphics[width=12cm]{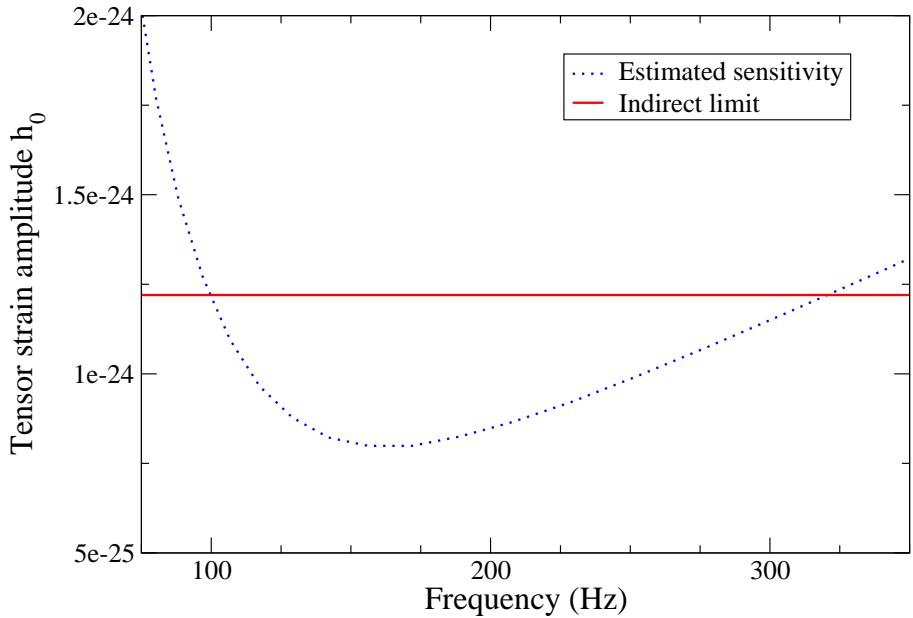}}
\caption{
\label{plot}
Estimated sensitivity of an S5 search compared to the indirect limit on GW
emission for Cas A.
Search parameters are the fiducial ones described in the text.
}
\end{figure}

Finally we estimate the computational cost and sensitivity of a directed search.
Preliminary runs on nodes of the APAC cluster \cite{APAC} find a timing of
about $6\times10^{-7}$~s per template per SFT.
Assuming 30-minute SFTs and two interferometers with 70\% duty cycle, the
computing time for the search (exclusive of Monte Carlo simulations to compute
upper limits) is
\begin{equation}
\mbox{20 days } \left( \frac{f_{\max}} {\mbox{300 Hz}} \right)^3 \left(
\frac{\mbox{300 years}}{\tau} \right)^3 \left( \frac{T_\mathrm{span}}{\mbox{12
days}} \right)^7 \left( \frac{200} {\mbox{nodes}} \right).
\end{equation}
For these fiducial parameters and two interferometers with the initial LIGO
design noise spectrum~\cite{LIGO:SRD} and 70\% duty cycle, the sensitivity
curve \eref{h95} is plotted in \fref{plot}.
The minimum of the curve (smallest detectable $h_0$) is
\begin{equation}
\hspace{-\mathindent}
8.0\times10^{-25} \left( \frac{\mbox{12 days}} {T_\mathrm{span}}
\right)^{-1/2}
\quad \mbox{or} \quad
8.0\times10^{-25} \left( \frac{f_{\max}} {\mbox{300 Hz}} \right)^{3/14}
\left( \frac{\mbox{300 years}}{\tau} \right)^{3/14},
\end{equation}
where the latter scalings allow $T_\mathrm{span}$ to vary at fixed
computational cost and are useful for evaluating searches for other objects.
Combining the previous two equations indicates that the sensitivity only
improves as the 14th root of the computational cost, and thus there is not
much point in integrating for significantly longer without an improved
semi-coherent analysis method.

Thus we see that this search for Cas A, when completed on S5 data, will beat
the fiducial indirect limit on GW emission from about 100 to 300~Hz.
This will double the number of objects (after the Crab pulsar) for which
initial LIGO has beaten an indirect limit.
Similar searches can be made for other suspected young neutron stars.

\ack
We are grateful to the LSC internal reviewers for helpful comments on the
manuscript.
This work was supported by Australian Research Council grant DP-0770426 and
USA National Science Foundation grants PHY-0245649, PHY-0555628, and
cooperative agreement PHY-0114375 (the Penn State Center for Gravitational
Wave Physics).
This article has been assigned LIGO document number P070123-02-Z.

\section*{References}
\bibliographystyle{unsrt}

\bibliography{CasAMethodsPaper}

\end{document}